\documentclass[12pt,preprint]{aastex}

\usepackage{wasysym}

\shorttitle{}
\shortauthors{Dimitri Veras}

\begin{document}
\baselineskip 14.pt

\title{Planetary Systems Around White Dwarfs}

\author{Dimitri Veras$^{1,2}$}
\affil{(1) Centre for Exoplanets and Habitability, University of Warwick, \\ 
Coventry CV4 7AL, United Kingdom}
\affil{(2) Department of Physics, University of Warwick, \\ 
Coventry CV4 7AL, United Kingdom}

\begin{abstract} 
White dwarf planetary science is a rapidly growing field of research featuring a diverse set of observations and theoretical explorations. Giant planets, minor planets, and debris discs have all been detected orbiting white dwarfs. The innards of broken-up minor planets are measured on an element-by-element basis, providing a unique probe of exoplanetary chemistry. Numerical simulations and analytical investigations trace the violent physical and dynamical history of these systems from au-scale distances to the immediate vicinity of the white dwarf, where minor planets are broken down into dust and gas and are accreted onto the white dwarf photosphere. Current and upcoming ground-based and space-based instruments are likely to further accelerate the pace of discoveries. 
\end{abstract}

\section{Introduction}

White dwarf exoplanetary systems provide windows into composition, physical processes and minor planets (asteroids, comets, moons, and interior fragments of larger planets) which are unavailable in main-sequence investigations. Almost every known exoplanet orbits a star that will become a white dwarf, and these exoplanets will either survive or be destroyed throughout the stellar transformation. En route, the exoplanets' perturbations on smaller bodies contribute to the latter's physical and dynamical evolution, allowing them to approach and be destroyed by the white dwarf. The resulting remnants are ubiquitously observable, such that the occurrence rate of white dwarf planetary systems is comparable to that of main-sequence exoplanetary systems.

In fact, planetary systems around white dwarfs have now been observed at a variety of stages of destruction, from fully intact planets to those which have been shredded down to their constituent chemical elements. Theoretical models need to explain an expanding variety of planetary system signatures, which include intact solid bodies, partially disrupting solid bodies, remnant discs and rings, dry, wet and differentiated chemical signatures, and accretion onto and diffusion within white dwarf photospheres. These observational signatures have been acquired from a striking diversity of discovery techniques and instruments, and new findings continue to surprise and delight. 

This article is focussed on the demographics of and theoretical explanations for white dwarf planetary systems, and does not highlight planetary system investigations which emphasize the effects of giant branch stars (the immediate precursors to white dwarf stars). For a pre-2016 review of giant branch planetary systems, see \cite{veras2016a}.

\begin{figure}
\centering
\includegraphics[width=1.0\textwidth]{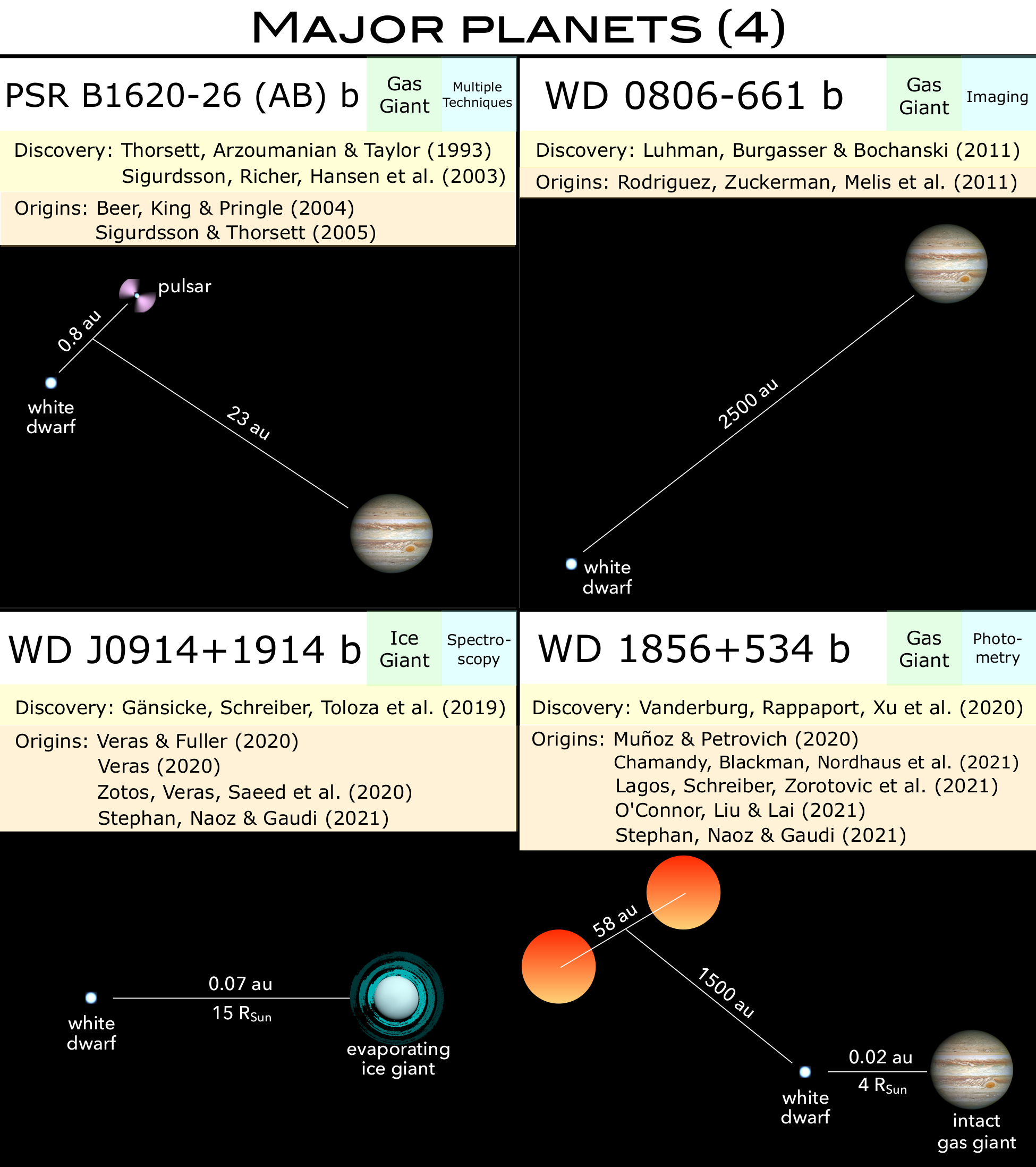}
\caption{Four major planets orbiting white dwarfs. Planet type is shown in green boxes, and the primary discovery method in blue boxes. The separations given are sky-projected, rather than actual.}
\label{Fig1}
\end{figure}

\section{Demographics}\label{demographics}

Signatures of white dwarf planetary systems can be split into four categories: (i) major planets (with radii $\gtrsim 10^3$ km), (ii) minor planets (with radii $\lesssim 10^3$ km), (iii) discs and rings, and (iv) photospheric metal chemistry (often characterized as ``pollution"). 

The raw numbers and occurrence rates of each of these four categories are dependent on one's confidence level about a particular observation or set of observations. An ongoing controversy in main-sequence exoplanetary science is whether to classify a potential planetary object as ``validated", ``confirmed" or as a ``candidate" or ``brown dwarf". In white dwarf planetary science, these terms are even less well-defined, particularly for the major and minor planets, partly because some spectroscopic discovery methods are new (whereas transit photometry and imaging are used similarly in both main-sequence and white dwarf communities). Given this caveat, this article will not use any of these quoted adjectives; readers can decide for themselves which are most appropriate through a careful analysis of the discovery papers. 

\begin{figure}
\centering
\includegraphics[width=1.0\textwidth]{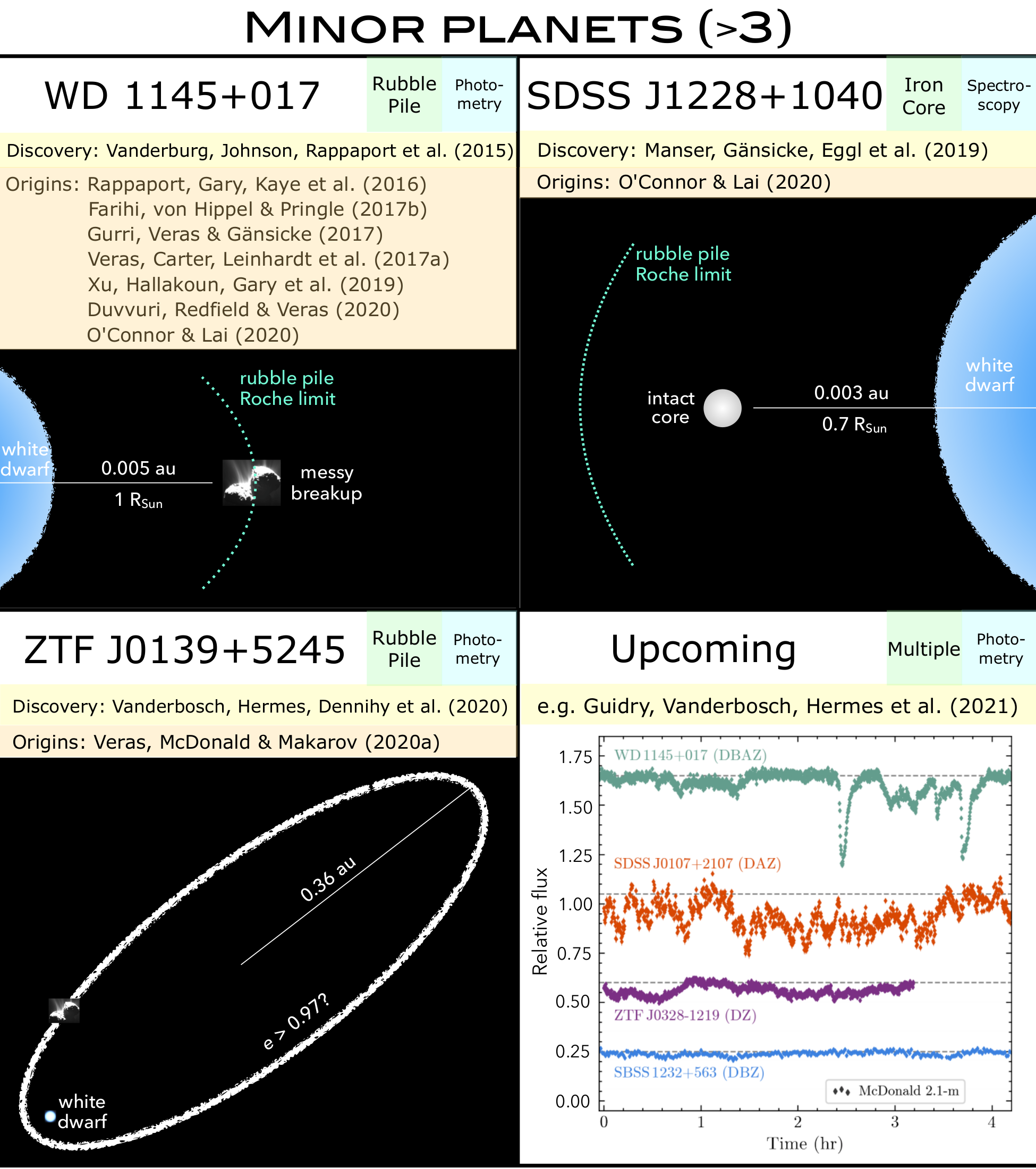}
\caption{Known and probable minor planets orbiting white dwarfs, at various early stages of disruption. The green boxes feature minor planet type, and the blue boxes feature primary discovery technique. The lower-right plot is © AAS. Reproduced with permission.}
\label{Fig2}
\end{figure}

In terms of raw numbers of white dwarf planetary systems, as of the beginning of 2021, this article assumes the discovery of 4 major planets, at least 3 minor planets, over 60 discs or rings, and over 1,000 white dwarfs with photospheric planetary debris (Figs. \ref{Fig1}-\ref{Fig4}). In terms of occurrence rates, about one-quarter to one-half of Solar neighbourhood white dwarfs feature photospheric metal chemistry. Discs are assumed to orbit nearly all these chemically enriched white dwarfs, and none of the un-enriched white dwarfs. Predicted occurrence rates of major and minor planets suffer from significant observational biases, and theoretical models could only fill in these gaps with multiple assumptions in the high degree-of-freedom parameter space (Fig. \ref{Fig6}). 

\subsection{Major planets}

The four major planets which orbit white dwarfs \citep{thoetal1993,sigetal2003,luhetal2011,ganetal2019,vanetal2020} are illustrated with cartoons in Fig. \ref{Fig1}. Although the sizes of these systems are not to scale, the placement of all four on the same figure highlights their diversity in terms of primary discovery method (imaging, spectroscopy, photometry, pulsar timing, astrometry) and architecture (circumbinary with a pulsar and white dwarf for the PSR B1620-26AB system, and circumstellar with a white dwarf for the other three, including two M-star companions in the WD~1856+534 system). The planet-white dwarf separations range from 0.02 au to about 2,500 au, with a notable absence of planets in the few au range due to observational bias. 

Significantly, all four planets are giant planets; so far, no known terrestrial-sized exoplanets around white dwarfs have been discovered, again likely due to observational bias. The mass of PSR B1620-26AB b is about $2.5 \pm 1.0 M_{\rm Jup}$ \citep{sigetal2003} and the mass of WD 0806-661 b is about $7 \pm 1 M_{\rm Jup}$ \citep{luhetal2011}. The masses of the other two planets are not as well constrained: for WD~J0914+1914~b, observations do not provide limits except for requiring that the planet is consistent with the mass of an ice giant \citep{ganetal2019}. Theoretical constraints, however, suggest that the planet is inflated and underdense \citep{verful2020}, which would restrict the potential ice giant mass range. The mass of WD~1856+534~b has an upper limit of $14M_{\rm Jup}$ \citep{vanetal2020}, and a lower limit of about $2M_{\rm Jup}$ \citep{aloetal2021}.

The probable origins of all four major planets are strikingly different. Because PSR B1620-26AB b is a cluster planet which orbits both a pulsar and a white dwarf, the dynamical history of the system likely includes a triple interaction within the cluster, potentially involving a common envelope \citep{beeetal2004,sigtho2005}. The wide separation of WD 0806-661 b ensures that it did not form in-situ \citep{rodetal2011}. Instead, the planet's current location my be explained by a previous gravitational scattering event within the system, or from a dissipative capture from an external companion or cluster. In contrast, the close separations of WD~J0914+1914~b and WD~1856+534~b require either (i) a gravitational scattering event during the white dwarf phase, plus some form of tidal shrinkage \citep{munpet2020,ocoetal2021,verful2020,veras2020,zotetal2020,steetal2021}, or (ii) survival of the planet within a common envelope of the white dwarf precursor \citep{chaetal2021,lagetal2021}.

\subsection{Minor planets}

Despite the crucial role of major planets as agents of debris delivery, they have been found in $<1\%$ of currently known white dwarf planetary systems \citep{dufetal2007,kleetal2013,kepetal2015,kepetal2016,couetal2019}. All other observations of signatures in these systems are thought to arise directly from minor planets. This subsection, which is based on Fig. \ref{Fig2}, summarizes the known minor planets which are intact or are in the early stages of disruption; subsections \ref{discs}-\ref{pollution} detail the more evolved stages of minor planet debris. 

Transit photometry has so far been responsible for all minor planet discoveries 
\citep{vanetal2015,vanderboschetal2020,guietal2021} except in the SDSS J1228+1040 system \citep{manetal2019}. Unlike the transit curves which are usually seen in main-sequence exoplanetary science, the transit curves for the minor planets orbiting WD~1145+017 and ZTF~J0139+5245 (and possibly systems like SDSS~J0107+2107, ZTF~J0328-1219, and SBSS~1232+563) (i) are not of solid bodies but rather predominantly dusty effluences, (ii) contain trackable features which can change on weekly, monthly and yearly timescales, and (iii) can have maximum transit depths exceeding 50\%. These minor planets are hence not intact, but are in various stages of ongoing disruption.

The WD~1145+017 system \citep{vanetal2015} has received more attention than any other white dwarf planetary system, with over 20 papers dedicated to follow-up observations; see \cite{vanrap2018} for a pre-2018 review. This extensive attention was likely due to a few factors: (i) The system contains the first minor or major planet that was discovered close to a single white dwarf, providing stark affirmation that planets can reach such close orbits and thereby removing any lingering doubts about the origin of the photospheric and disc metals, (ii) the system is easily observed by professionals and amateurs alike due to a 4.5-hour orbital period of transiting debris, and (iii) the system is exciting and fun to observe because of the variable debris features.

Substantial theoretical work on this system \citep{rapetal2016,faretal2017b,guretal2017,veretal2017a,xuetal2019a,duvetal2020,ocolai2020} has placed constraints on the progenitor minor planet mass (about 10\% of Ceres' mass), bulk density (Vesta-like), interior structure (probably differentiated), placement relative to the circumstellar dust and gas in the system, and dynamical history through tidal shrinkage and circularization and ram pressure drag. Because the minor planet effluences orbit WD~1145+017 at a near-constant separation of $0.005$ au, which corresponds to the rubble-pile disruption (Roche) limit, the disruption in that system is assumed to be tidal.

\begin{figure}
\centering
\includegraphics[width=1.0\textwidth]{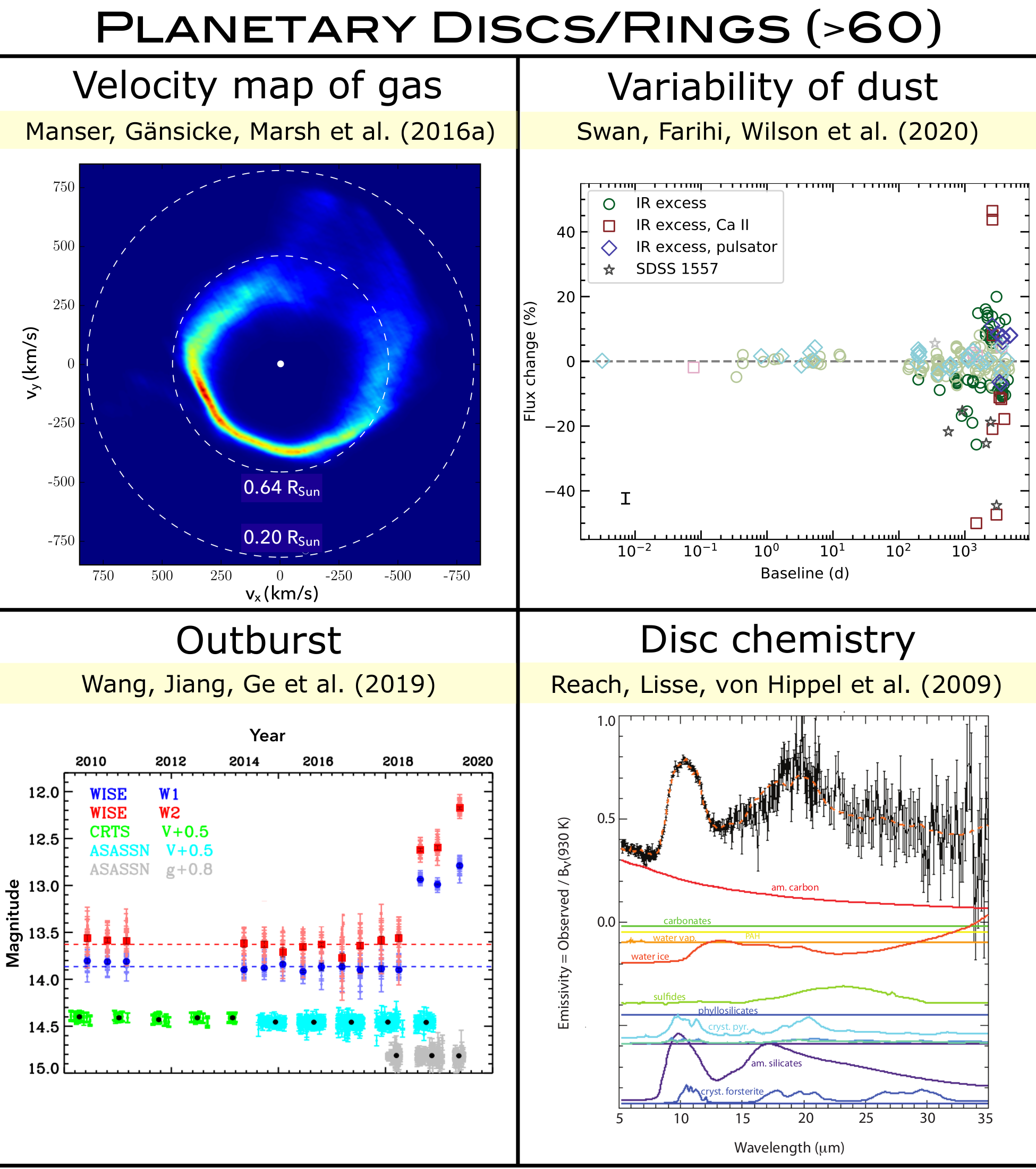}
\caption{Four highlighted aspects of planetary discs and rings orbiting white dwarfs. The velocity map ({\it upper left}) illustrates non-axisymmetric disc geometry. The dust variability plot ({\it upper right}) suggests widespread dynamical activity over yearly and decadal timescales, whereas the outburst event ({\it lower left}) is sudden. The chemistry plot ({\it lower right}) is our best example of dusty disc chemistry. The bottom two plots are © AAS. Reproduced with permission.
}
\label{Fig3}
\end{figure}

The origin of the disruption in ZTF~J0139+5245 \citep{vanderboschetal2020} is not as clear. The dusty effluences are on a 0.36 au orbit, which would need to have an eccentricity exceeding about 0.97 in order for tidal disruption to generate the debris; alternatively, rotational fission may occur outside of the Roche radius, allowing for breakup to occur beyond distances of 0.005 au \citep{veretal2020a}. Transit signatures of other minor planets \citep{guietal2021} represent, at this time, robust hints of minor planets rather than definitive detections.  

Unlike all of the minor planets detected by transit photometry, the spectroscopically-determined minor planet orbiting SDSS J1228+1040 \citep{manetal2019} was identified through variability in Ca II triplet emission, which provides geometric disc information. This minor planet is embedded within the disc, does not appear to be disrupting, and resides well within the rubble-pile Roche radius, at a distance of about 0.003 au. This combination of features suggests that the body is actually an iron-rich planetary core with tensile strength and internal viscosity \citep{ocolai2020}.

\subsection{Discs and rings}\label{discs}

The products of minor planet break-up have been commonly denoted in the literature as a ``disc" and are detected through infrared excesses on spectral energy distributions (for dust) and key emission and absorption spectral signatures (for gas). This disc nomenclature refers to structures contained within about 0.005 au ($1R_{\rm Sun}$) of the white dwarf and ignores (unobservable) debris fields or belts extending beyond about 10 solar radii from the white dwarf. This nomenclature may also erroneously give the impression of an ordered, circular-shaped structure with well-defined inner and outer boundaries. In fact, these structures can be wispy, ring-like, eccentric \citep{ganetal2006,denetal2016,manetal2016a,nixetal2021} and variable (Fig. \ref{Fig3}). 

The number of detections of these discs and rings now exceed about 60 \citep{denetal2020,manetal2020,meletal2020,xuetal2020,genetal2021} and their discoveries extend back to \cite{zucbec1987}; for a pre-2016 review, see \cite{farihi2016}, and for a more recent review in comparison to main-sequence debris discs, see \cite{cheetal2020}. All of these discs are circumstellar except for one circumbinary exception \citep{faretal2017a}. 

Nearly all of the discs contain dust; a notable exception being the all-gas disc around WD~J0914+1914, which also hosts a major planet (Fig. \ref{Fig1}). About 1-3\% of all white dwarfs contain observable dusty discs \citep{faretal2009} and about 4\% of dusty discs contain observable gas \citep{manetal2020}; a couple of notable discs which contain both observable gas and dust are WD~1145+017 and SDSS~J1228+1040, which also host minor planets (Fig. \ref{Fig2}). The true fraction of white dwarfs with discs is likely much higher \citep{rocetal2015,bonetal2017}, and probably is similar to the fraction of white dwarfs containing photospheric metal debris (25-50\%). White dwarf discs do not appear to be more numerous in systems with binary main-sequence stellar companions \citep{wiletal2019}.

\begin{figure}
\centering
\includegraphics[width=1.0\textwidth]{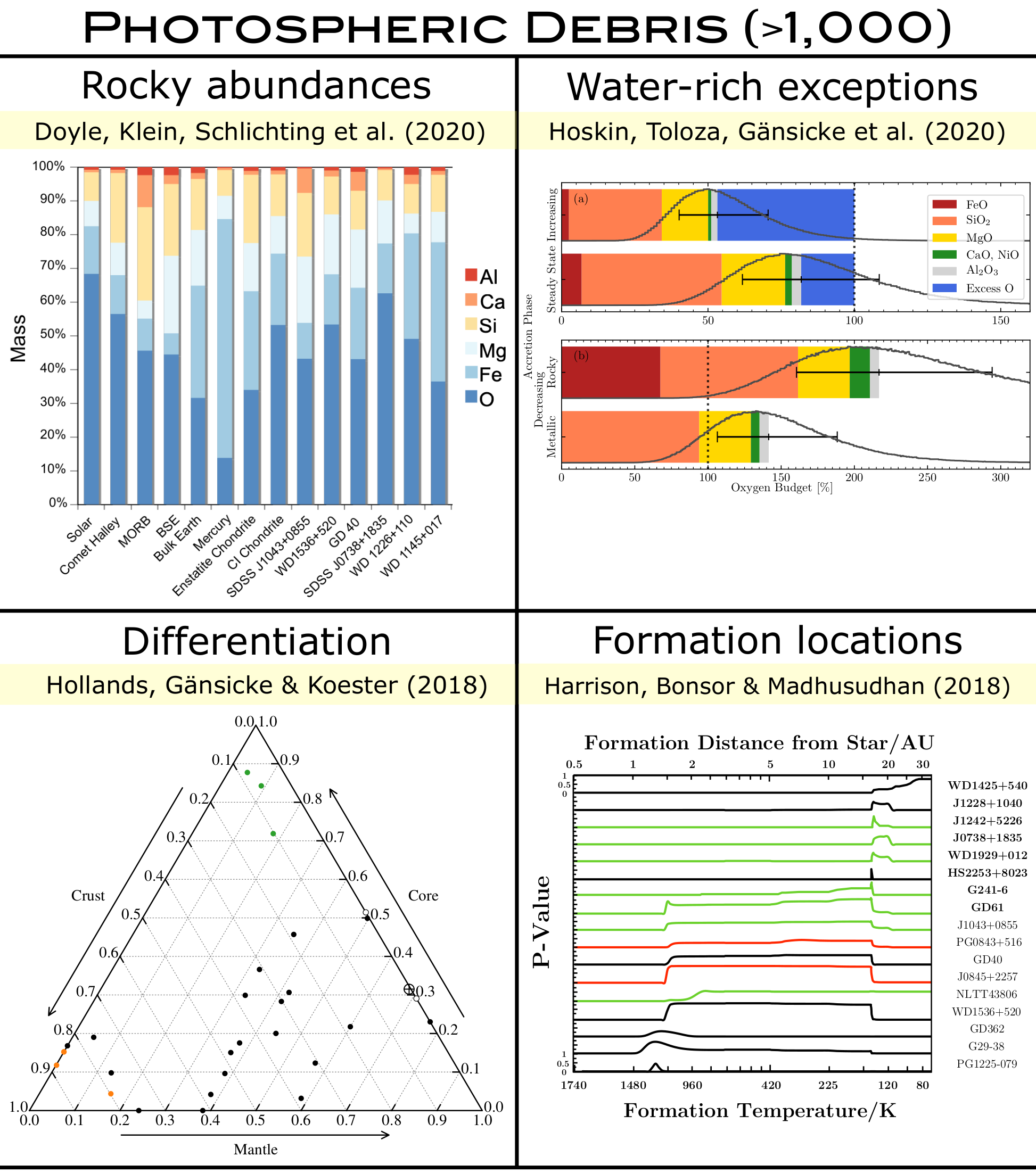}
\caption{Four examples of exoplanetary chemistry from debris in white dwarf photospheres, commonly referred to as white dwarf pollution. The majority of pollutants are dry and Earth-like ({\it upper left}) despite notable water-rich exceptions ({\it upper right}). Differentiated progenitors of the pollutants ({\it lower left}) may be traced back to formation locations in the protoplanetary birth disc ({\it lower right}). The upper left plot is © AAS. Reproduced with permission.}
\label{Fig4}
\end{figure}

\begin{figure}
\centering
\includegraphics[width=0.8\textwidth]{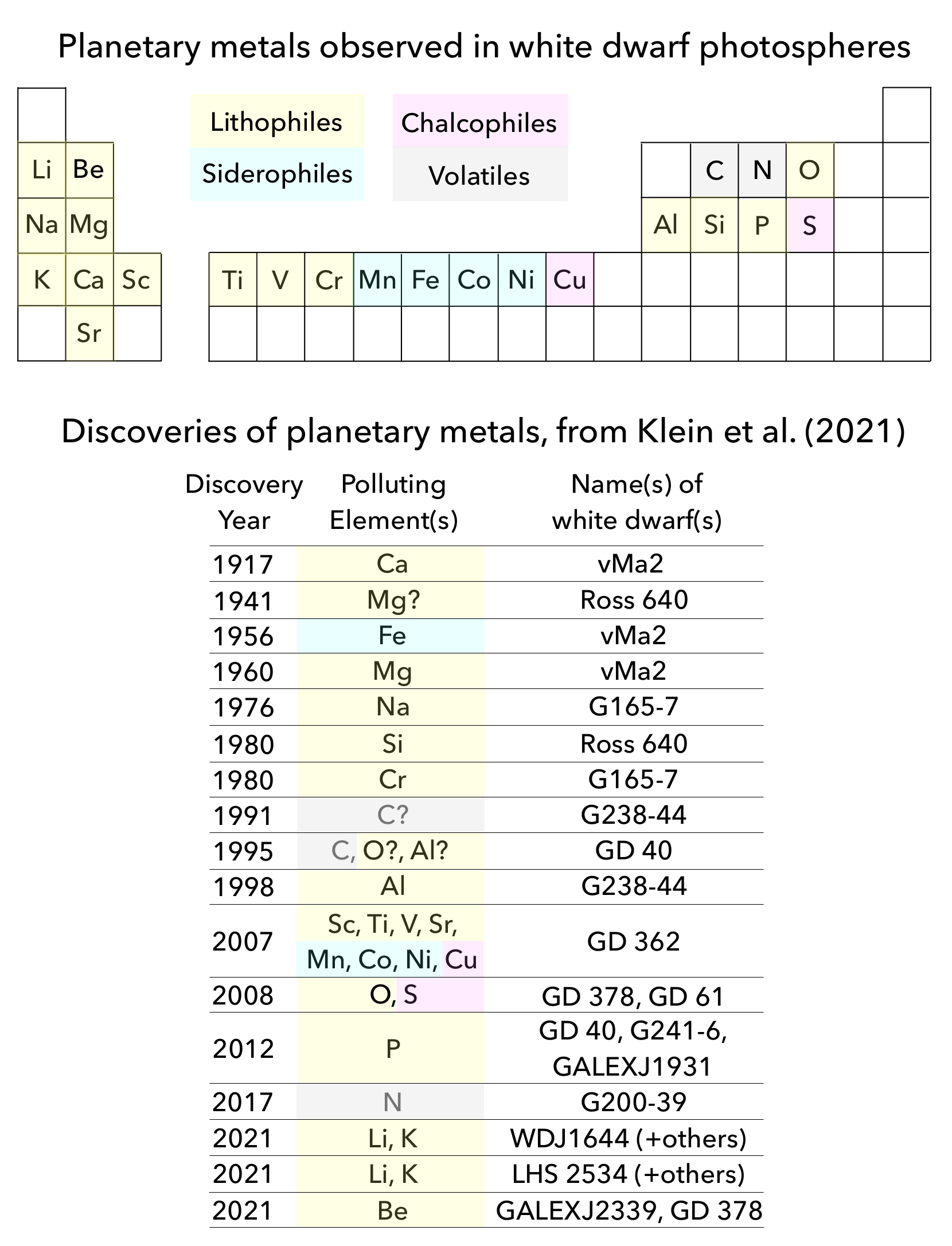}
\caption{All planetary metals (chemical elements heavier than helium) found in white dwarf photospheres, partitioned by the geochemical Goldschmidt classification. Hydrogen is also accreted and can sometimes be measured. The table of discoveries is taken from the first three columns of Table 1 of \cite{kleetal2021}; for notes, references and additional details, see that paper.}
\label{Fig5}
\end{figure}

One of the most exciting aspects of these discs is that they showcase activity beyond standard Keplerian motion around the white dwarf. One manifestation of the activity is through flux changes due to some physical process in the dust \citep{xujur2014,faretal2018b,xuetal2018,wanetal2019,rogetal2020,swaetal2020,wiletal2020}. Another manifestation is through the secular (long-term) precession of gas, as well as its variability \citep{wiletal2014,manetal2016a,manetal2016b,manetal2019,manetal2021,redetal2017,cauetal2018,denetal2018,denetal2020,foretal2020,genetal2021}.

Both dust and gas can also represent windows into composition, although planetary chemistry is primarily obtained from photospheric debris measurements (subsection \ref{pollution}). Pre-JWST, the only dusty disc close enough and bright enough to Earth for which chemical constraints can be modelled in detail orbits G~29-38 \citep{reaetal2005,reaetal2009}; broader chemical dust models can still be fit around other white dwarfs \citep{xuetal2018}. Circumstellar gas offers greater opportunities \citep{meletal2010,haretal2011,haretal2014,xuetal2016} and crucially allows one to match the chemistry in the disc with the chemistry in the photosphere \citep{ganetal2019,steele2021}.

\subsection{Photospheric chemistry}\label{pollution}

After being ground down and converted into gas, the planetary debris is finally accreted onto the photosphere of the white dwarf. Particularly strong objects such as SDSS J1228+1040~b may also impact the white dwarf photosphere directly when dynamically perturbed. The high density of the white dwarf then stratifies the accreted debris into its constituent chemical elements. These so-called ``metals" are detectable because they stand out against a backdrop of primarily hydrogen and helium, giving rise to the common term ``pollution". Uniform samples of single white dwarfs with similar properties have revealed that 25-50\% of all single white dwarfs are polluted \citep{zucetal2003,zucetal2010,koeetal2014}, with the total number standing at over 1,000 \citep{couetal2019}. For white dwarfs in binaries \citep{zuckerman2014,bonetal2021}, winds from main-sequence stellar companions negligibly contribute to photospheric pollution unless the binary orbit is smaller than a few au \citep{debes2006,veretal2018b}.

The accreted matter directly probes exoplanetary chemistry, and at a high level of detail (Figs. \ref{Fig4}-\ref{Fig5}). The pioneering study of \cite{zucetal2007} demonstrated how abundances of over a dozen metals in a single white dwarf can be pieced together to infer the chemical composition of the progenitor minor planet. Since then, a multitude of abundance studies have focussed on interesting individual white dwarfs or ensembles of white dwarfs; see \cite{juryou2014} for a pre-2014 review. Usually, these studies \citep[e.g.][]{kleetal2010,ganetal2012,xuetal2013,baretal2014,juretal2014,xuetal2014,juretal2015,wiletal2015,faretal2016,melduf2017,xuetal2019b} analyze white dwarfs with at least three different metals; the majority of polluted white dwarfs have only one or two (typically calcium and magnesium) which exceed the detection threshold. 

The high quality of the data and modelling now allows for detailed comparisons to the compositions of a variety of solar system meteorites and planets \citep{bloetal2019,swaetal2019,doyetal2019,doyetal2020}, including investigations of the carbon-to-oxygen ratio \citep{wiletal2016}, the level of differentiation in and mixing amongst the progenitor minor planets \citep{juretal2013,holetal2018,bonetal2020,turwya2020}, and the formation locations and level of post-nebula de-volatilization of the progenitors \citep{haretal2018,haretal2021}. Although the vast majority of pollutants are ``dry" (volatile-poor) and Earth-like in composition, exciting exceptions include white dwarfs with water-rich progenitors \citep{faretal2013,radetal2015,genetal2017,hosetal2020,izqetal2021}, a Kuiper belt-like comet/asteroid \citep{xuetal2017} and an exomoon \citep{doyetal2021}. Over 20 metals in total have now been found (Fig. \ref{Fig5}), with the last four being nitrogen \citep{xuetal2017}, lithium and potassium \citep{holetal2021,kaietal2021}, and beryllium \citep{kleetal2021}.

\begin{figure}
\centering
\includegraphics[width=1.0\textwidth]{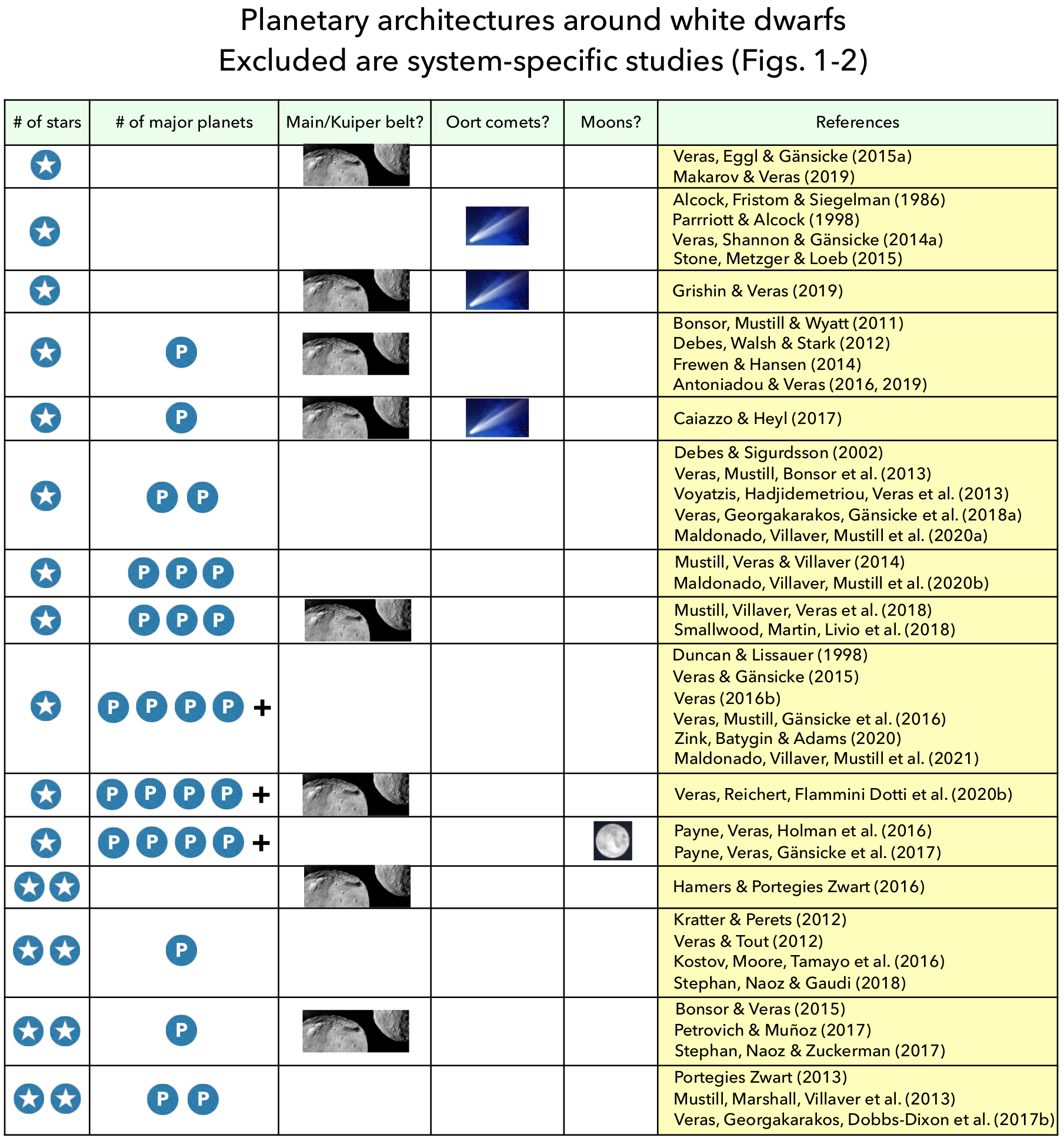}
\caption{A rough breakdown of parameter space for investigations of planetary architectures around white dwarfs.}
\label{Fig6}
\end{figure}
 
\section{Explanations}

By itself, the drive to understand the fate of planetary systems has prompted innovative theoretical investigations. Additional motivation arises from the plethora of increasingly diverse observations of white dwarf planetary systems that were summarized in Section \ref{demographics}. These theoretical investigations can be split into two categories, which are the subject of subsections \ref{arch} and \ref{close}: (i) planetary architectures and (ii) activity close to and inside of the white dwarf. A comprehensive accounting of theoretical investigations pre-2016 can be found in \cite{veras2016a}.

\subsection{Planetary architectures}\label{arch}

All planetary signatures in single white dwarf systems are the end result of gravitational and radiative perturbations amongst major and minor planets at distances well beyond several solar radii, usually beyond observable limits. Because these perturbations are linked with the evolution of the central star, investigations often evolve both the star and planetary system simultaneously. 

The parameter space to explore is large: at minimum, 6 orbital elements and a mass must be provided for each major and minor planet. Further, observations of these planets are not yet numerous enough to constrain this parameter space. Consequently, investigators have largely relied on a ``divide and conquer" approach in the literature, restricting individual investigations to a specific family of architectures. These families are broadly categorized in Figure \ref{Fig6} in terms of number of stars, number of major planets, and whether minor planets were included in each investigation. 

Amongst minor planets, different types have been modelled: (i) most commonly, asteroids in analogues of the main belt, Kuiper belt and scattered disc, (ii) comets in Oort cloud analogues \citep{alcetal1986,paralc1998,veretal2014a,stoetal2015,caihey2017,griver2019}, and (iii) exo-moons \citep{payetal2016,payetal2017}. How much each class of object contributes to white dwarf pollutants remains debatable, even though the term ``asteroids" has conventionally been adopted as a catch-all for minor planets. By analogy with the solar system, the total mass in moons can easily exceed the total mass in a main belt analogue, and the total mass of Oort cloud comets is uncertain by many orders of magnitude. Chemically, although dry asteroids and moons provide the best match to the majority of white dwarf pollutants, the mounting number of volatile-rich minor planet progenitors (subsection \ref{pollution}) suggests that relying on dry asteroids alone to represent pollutants is incorrect.

\begin{figure}
\centering
\includegraphics[width=1.0\textwidth]{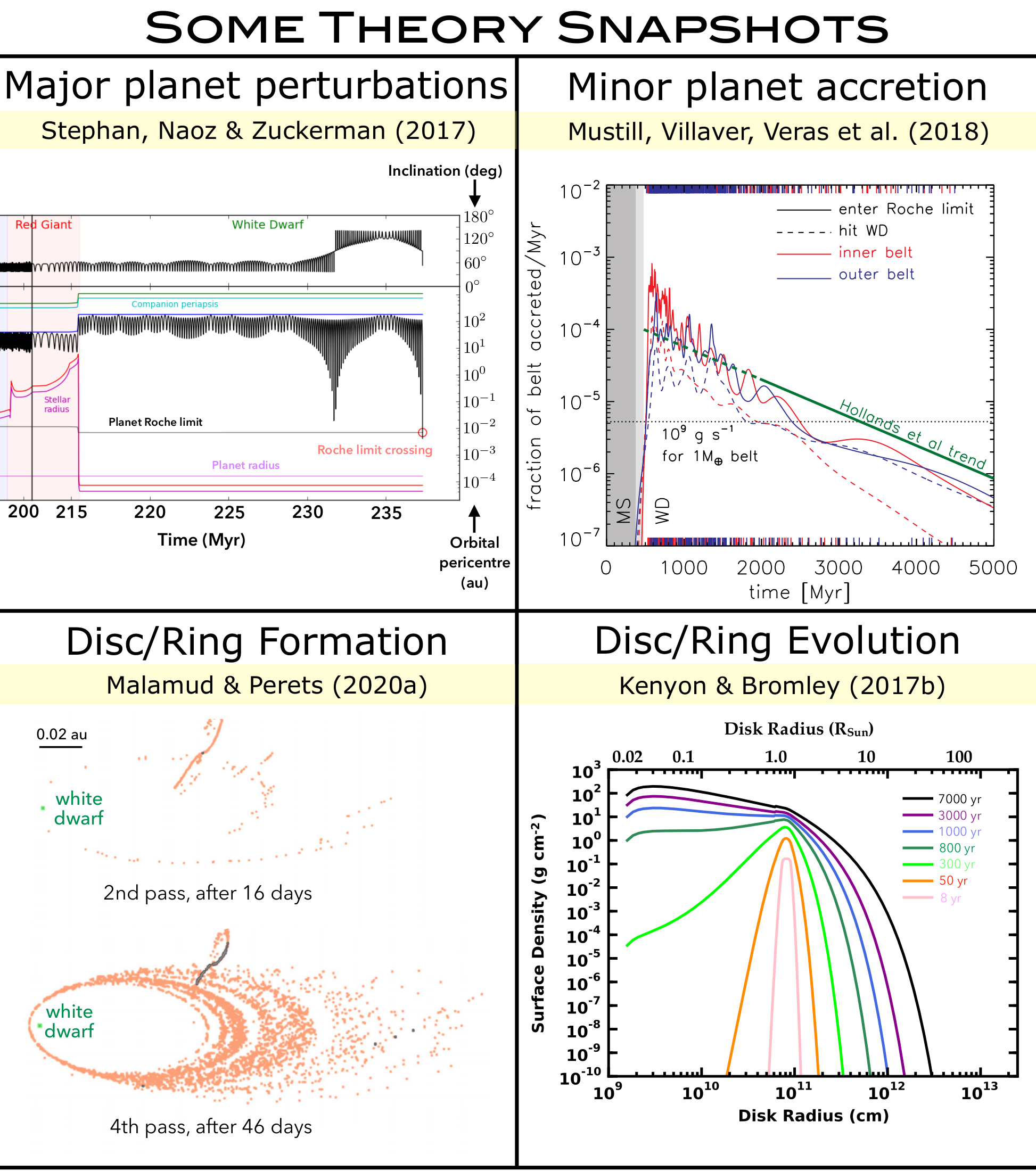}
\caption{Illustrative examples of planetary architecture evolution for an exo-Neptune entering the white dwarf Roche radius {\it (upper left)}, exo-asteroid belts accreting onto a white dwarf in a multi-planet system {\it (upper right)}, disc formation {\it (lower left)} and disc evolution {\it (lower right)}. The upper left and lower rights plots are © AAS. Reproduced with permission.}
\label{Fig7}
\end{figure}
 
Architecture studies with zero major planets typically feature physics other than just point-mass $N$-body gravitational interactions, including sublimation of volatile patches on active asteroids \citep{veretal2015a}, rotational fission of asteroids due to spin angular momentum exchange \citep{makver2019}, the effect of Galactic tides and stellar flybys on comets \citep{alcetal1986,paralc1998,veretal2014a,stoetal2015}, capture of asteroids and comets within a gaseous disc \citep{griver2019}, and von Zeipel-Lidov-Kozai perturbations due to a stellar companion \citep{hampor2016,steetal2017}. 

Although these mechanisms can succeed in polluting the white dwarf and populating an extant disc, at least one major planet \citep{bonetal2011,debetal2012,frehan2014,antver2016,antver2019,caihey2017} greatly facilitates dynamical delivery, particularly when a binary stellar companion is involved \citep{bonver2015,petmun2017,steetal2017,steetal2018} (see Fig. \ref{Fig7}). Circumbinary studies with one major planet \citep{kraper2012,vertou2012,kosetal2016} have yet to feature minor planets.

Investigations with two major planets \citep{debsig2002,portegieszwart2013,musetal2013,veretal2013,voyetal2013,veretal2017b,veretal2018a} or more allow for the possibility of (i) determining the fate of known main-sequence multi-planet exosystems, even if re-scaled \citep{maletal2020a,maletal2020b,maletal2021} and (ii) dynamical activity due to gravitational instability amongst the planets themselves. Depending on the initial configuration, the instability between the two planets can occur at any time during the white dwarf phase. However, with two planets, only one instance of instability is possible, and usually eliminates one planet from the system through escape or collision with the white dwarf. 

In contrast, studies with three major planets allow for multiple generations of instabilities \citep{musetal2014,musetal2018,maletal2020b} and secular resonance shifts due to engulfment of one of the planets \citep{smaetal2018}. With four or five major planets, investigations can model outer solar system analogues around white dwarfs, including the future Sun \citep{dunlis1998,veras2016b,veretal2020b,zinetal2020}. As the number of major planets increases further \citep{vergan2015,veretal2016,maletal2021}, an emerging trend is that their tendency to linger and meander after gravitational instabilities increases with (i) total number of planets, (ii) decreasing planet mass, and (iii) increasing orbital eccentricities. Such meandering can activate previously dynamically stagnant reservoirs of minor planets, potentially enhancing pollution and disc formation prospects.

\begin{figure}
\centering
\includegraphics[width=1.0\textwidth]{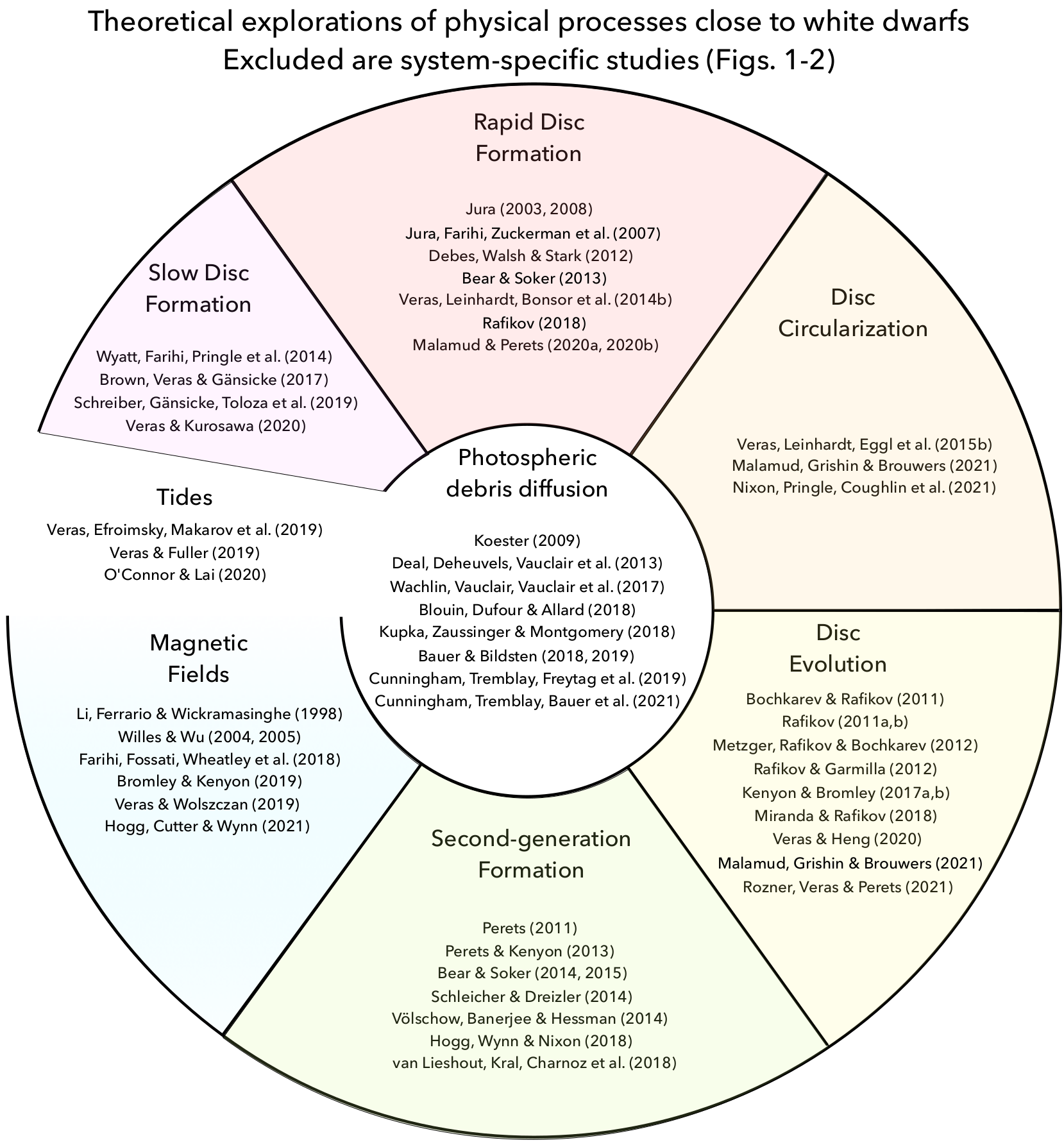}
\caption{Breakdown of the literature exploring different physics close to and inside of the white dwarf; shaded regions indicate the involvement of a disc or ring.}
\label{Fig8}
\end{figure}

\subsection{Physics close to the white dwarf}\label{close}

An important end product of the dynamical evolution of planetary architectures during the white dwarf phase is matter (planets, moons, comets, asteroids, dust) which veers close (within several Solar radii) to the white dwarf. At this stage, several different physical effects outlined in Fig. \ref{Fig8} become important; the shaded regions usually involve discs or rings. This subsection will discuss these effects in a clockwise direction from the illustration, starting with tides.

All objects approaching a white dwarf will experience gravitational tides, and these can alter the orbital eccentricity and semimajor axis of large asteroids and major planets whose pericentre distances exceed the Roche radius. The tidal interaction between white dwarfs and solid bodies \citep{veretal2019,ocolai2020} and gaseous bodies \citep{verful2019} reveal non-trivial dependencies on the internal structure and spin of the object. Extant discs or rings produced from minor planet destruction are unlikely to be massive enough to play a role in the tidal interaction. Hence, the presence and movement of a major planet by itself would not necessarily imply the existence of a disc or ring.

These discs and rings can be formed by a variety of mechanisms. Slow methods of forming a disc include the continuous or stochastic destruction of a collection of minor planets over time \citep{wyaetal2014,broetal2017}, the gradual accumulation of crater impact ejecta \citep{verkur2020} and the gradual accumulation of evaporated planetary atmospheres \citep{schetal2019}. The one rapid formation mechanism is the tidal destruction of one large object representing either a large asteroid or major planet \citep{jura2003,jura2008,juretal2007,debetal2012,beasok2013,veretal2014b,rafikov2018,malper2020a,malper2020b}. Potential byproducts of rapid formation is the ejection of some minor planets from the system \citep{rafikov2018,malper2020b}, the development of significant substructure within the newly formed disc \citep{malper2020a,malper2020b} (see Fig. \ref{Fig7}), and a size distribution of remnants ranging from dust to asteroids.

Disc formation mechanisms relying on tidal destruction typically assume that the initial orbit of the tidally destroyed object is highly eccentric; in order to survive the giant branch phases of evolution, the object must arrive at the white dwarf at a distance of at least a few au. This idea has been observationally reinforced through the likely eccentric debris orbiting ZTF J0139+5245 \citep{vanderboschetal2020}. This debris might be in the process of circularizing. Fittingly, the process of and timescale for circularizing debris has become a subject in its own right \citep{veretal2015b,malamud2021,nixetal2021}.

The evolution of the resulting roughly circular disc or ring \citep{bocraf2011,rafikov2011a,rafikov2011b,metetal2012,rafgar2012,kenbro2017a,kenbro2017b,mirraf2018,verhen2020,malamud2021,rozetal2021} incorporates a variety of physics, partly because gas generated from sublimation or collisions mixes with the dust. Given that analytic solutions for the time evolution of this gas-dust mixture are limited, employment of numerical codes would be required for more detailed modelling (see Fig. \ref{Fig7} for an example evolution), especially for an individual known exosystem.

For single white dwarfs, speculation that ``second-generation" planets could form from disc material \citep{beasok2015} was addressed by \cite{vanetal2018}. The latter demonstrated that coagulation of second-generation minor planets is possible only in relatively massive discs with masses comparable to Io or Mercury. Formation of such massive discs is assumed to be rare, particularly because of the theoretical constraints on the masses of the minor planets in the WD~1145+017, SDSS~J1228+1040 and ZTF~J0139+5245 systems (see the references within Fig. \ref{Fig2}). However, for white dwarfs in binary systems, both second-generation discs \citep{perets2011,perken2013,hogetal2018} as well as second-generation planets \citep{beasok2014,schdre2014,voletal2014} appear to be more plausibly common.

Magnetic fields, if present, could play a role in much of the physics that was already mentioned in this subsection. Observationally, approximately 20\% of white dwarfs have kG-scale or larger magnetic fields, and about 10\% have fields stronger than 1 MG \citep{feretal2015,holetal2015,lanbag2019}. The magnetic fields in these white dwarfs could affect the evolution of major planets \citep{lietal1998,wilwu2004,wilwu2005,verwol2019}, minor planets \citep{broken2019} and disc material \citep{faretal2018a,hogetal2021}, and might open up a new avenue for the discovery of planetary remnants \citep{ganetal2020}. 

The fate of all close material smaller than major planets will be accretion onto the white dwarf. In the photosphere, the sinking timescales and diffusion properties of the planetary debris crucially affect the modelling of planetary chemistry, including whether the accretion is in a ``steady-state", ``increasing" or ``decreasing" phase (see upper-right panel of Fig. \ref{Fig4}). Each chemical element has a different sinking timescale \citep{koester2009,bloetal2018}, and is affected by thermohaline mixing \citep{deaetal2013,wacetal2017,baubil2018}, overshooting \citep{kupetal2018,baubil2019,cunetal2019} and horizontal, as well as vertical, sedimentation \citep{cunetal2021}.

\begin{figure}[t]
\centering
\includegraphics[width=1.0\textwidth]{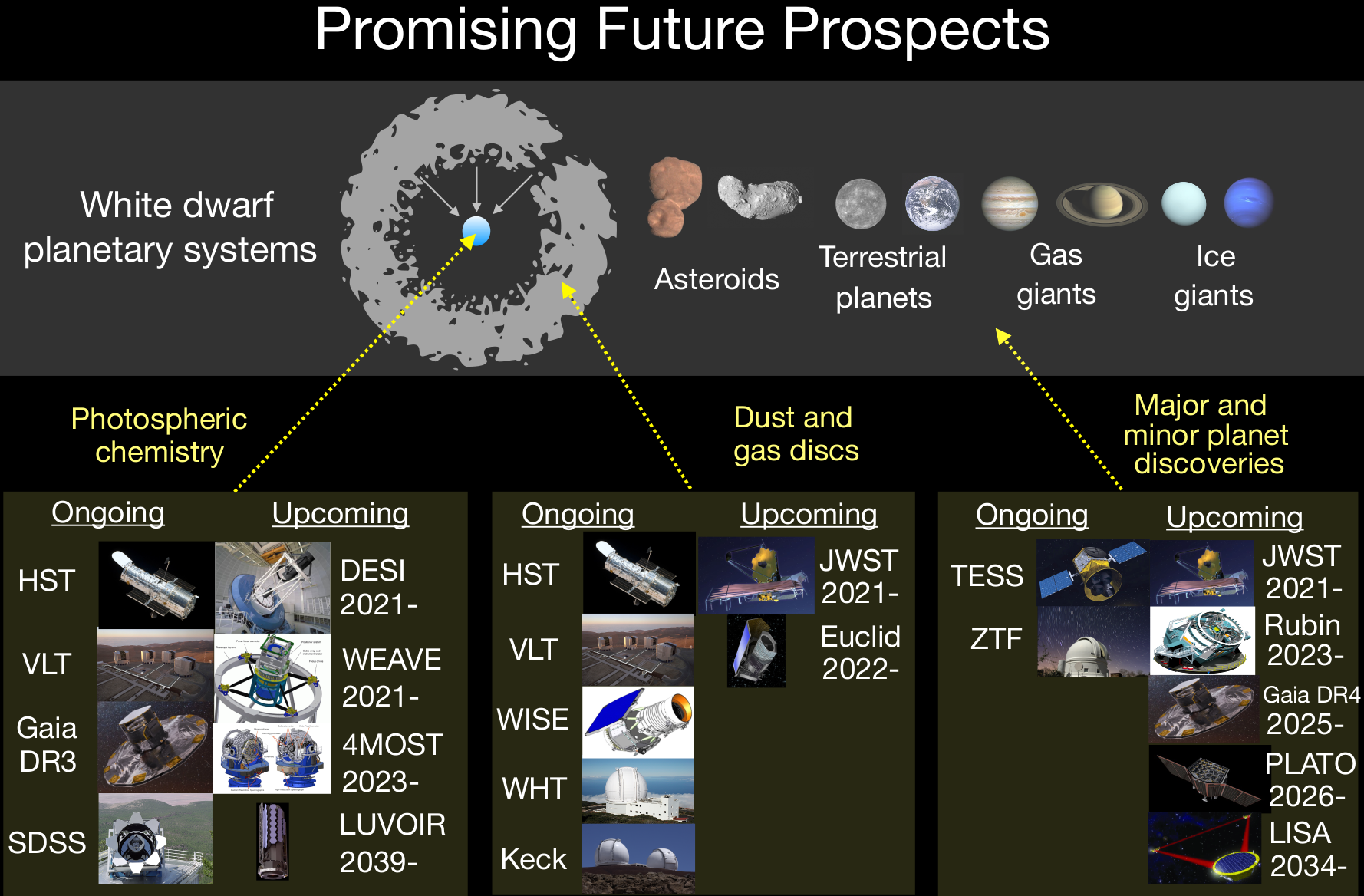}
\caption{A representative set of ongoing and upcoming ground-based and space-based facilities which represent the observational future of white dwarf exoplanetary science.}
\label{Fig9}
\end{figure}

\begin{figure}[h]
\centering
\includegraphics[width=1.0\textwidth]{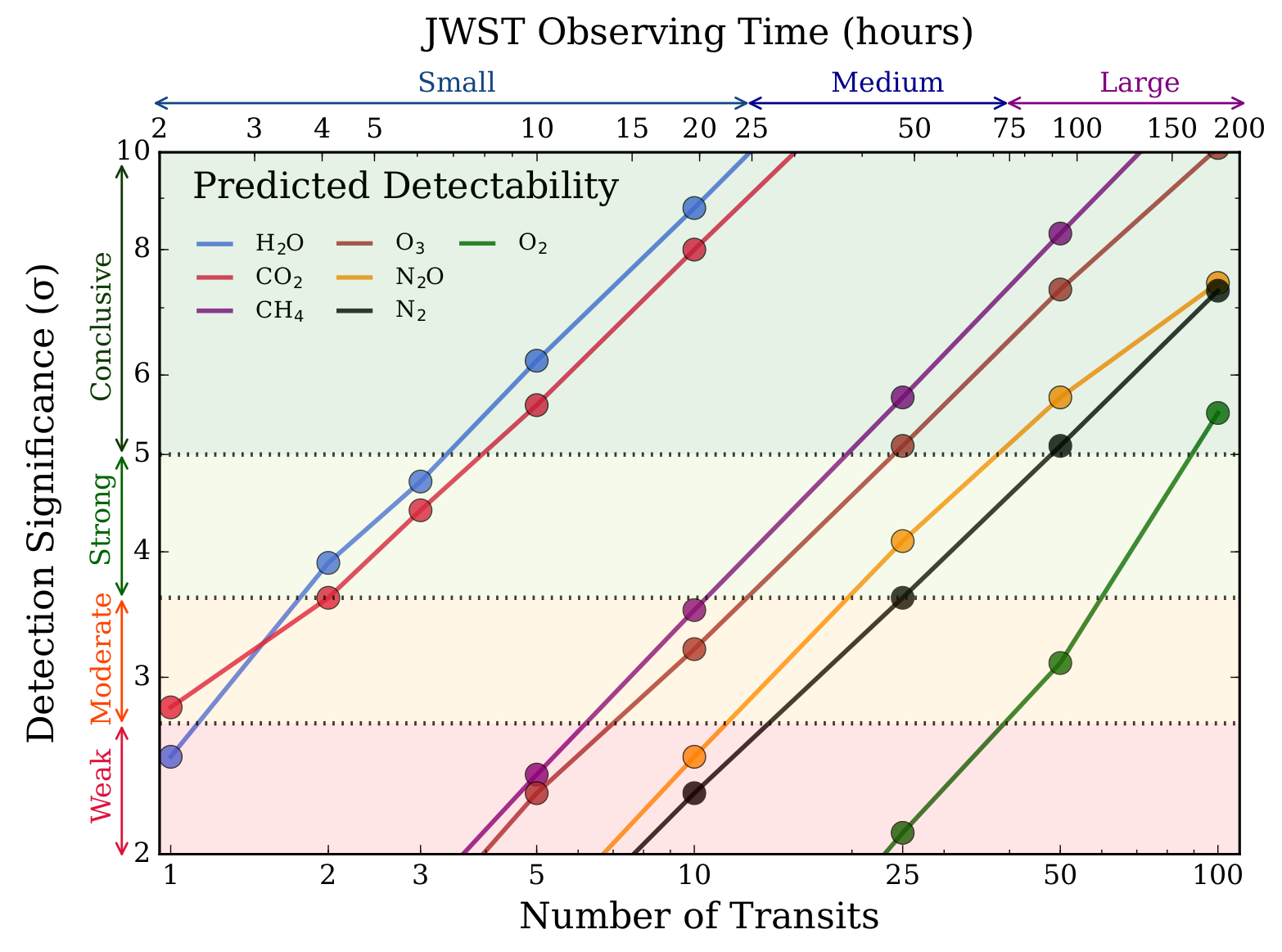}
\caption{The required number of JWST transits to detect molecular species in the atmosphere of an Earth-like planet orbiting a white dwarf. From \cite{kaletal2020}. This plot is © AAS. Reproduced with permission.}
\label{Fig10}
\end{figure}

\section{The future}

White dwarf planetary science is a rapidly growing field; only one of the references in Figs. \ref{Fig2}-\ref{Fig4} predates 2015, and about half of all the references in Figs. \ref{Fig6} and \ref{Fig8} were written after 2015. Prospects for future discoveries are summarized in Fig. \ref{Fig9}. A recent eight-fold increase in the total population of white dwarfs \citep{genetal2019} will help multiple ongoing and future facilities to continue characterizing photospheric chemistry, detecting dust and gas in discs and rings \citep{fanetal2020}, and discovering additional minor and major planets \citep{peretal2014,corkip2019,danetal2019,tamdan2019}. Such advances will motivate a variety of theoretical studies, such as those which incorporate sophisticated global hydrodynamical simulations of circumstellar gas and dust, planet formation around the highest mass progenitor host stars \citep{veretal2020c}, the mapping of orbital architectures with nebular chemistry and exo-meteorite families \citep{haretal2021}, and even tracking the prevalence of plate tectonics through stellar evolution \citep{juretal2014}.

Finally, by analogy with main-sequence exoplanetary systems, one other area of increasing interest is planetary atmospheric chemistry \citep{ganetal2019} and the potential for habitability around white dwarfs \citep{agol2011,fosetal2012,barhel2013,ramkal2016,kozetal2018,kozetal2020}. In fact, the relatively large radius ratio between a major planet and a white dwarf would allow for exquisite atmospheric characterization of terrestrial planets with JWST (Fig. \ref{Fig10}) \citep{kaletal2020}, and dynamical studies of planetary architectures (Fig. \ref{Fig6}) predict that such planets should exist.

\acknowledgements

\subsection*{Acknowledgements}
I thank my referees and Siyi Xu for thoroughly reading and commenting sagely on the entire manuscript. I have also received helpful and insightful feedback from Amy Bonsor, Tim Cunningham, Boris G\"{a}nsicke, Scott Kenyon, Beth Klein, Uri Malamud, Chris Manser, Tom Marsh, Alex Mustill, Steven Parsons, Alexander Stephan, Katja Stock, Silvia Toonen, Pier-Emmanuel Tremblay, and Andrew Vanderburg. I gratefully acknowledge the support of the STFC via an Ernest Rutherford Fellowship (grant ST/P003850/1).

\end{document}